\documentclass[twocolumn,english,aps,pra]{revtex4}
\usepackage[T1]{fontenc}
\usepackage{amsmath,graphicx,amssymb,epsfig,babel,dsfont}
\usepackage{mathrsfs}
\usepackage{color}
\renewcommand{\Im}{{\rm Im}}
\newcommand{\Tr}{{\rm Tr}}
\newcommand{\rd}{{\rm d}}
\newcommand{\kb}{k_{\rm B}}

\newcommand{\ri}{{\rm i}}

\newcommand{\re}{{\rm e}}

\begin{document}

\title{Efficiency and Mechanism of heat flux rectification with non-reciprocal surface waves in Weyl-Semi-Metals}

\author{A. Naeimi}
\affiliation{Institut f\"{u}r Physik, Carl von Ossietzky Universit\"{a}t, 26111, Oldenburg, Germany}

\author{S.-A. Biehs}
\affiliation{Institut f\"{u}r Physik, Carl von Ossietzky Universit\"{a}t, 26111, Oldenburg, Germany}

\date{\today}

\begin{abstract}
	We reinvestigate the mechanism of near-field heat transfer rectification between two Weyl semimetal nanoparticles and a planar Weyl semimetal substrate via the coupling to non-reciprocal surface modes. We first show that the previously predicted rectification ratio of 2673 is incorrect and should rather be 1502. Furthermore we show that depending on the distance between the nanoparticles there can be a much more efficient heat flux rectification with ratios of about 6000. Furthermore, we identify a previously overlooked range of forward rectification and a range of strong backward rectification with rectification ratios larger than 8000 for relatively small Weyl node separations. We investigate the mechanism behind this large heat flux rectification and study its sensitivity with respect to certain material parameters and temperature showing that even larger rectification ratios up to 15000 are possible highlighting that certain Weyl semimetals are strong candidates for highly efficient heat flux rectification.
\end{abstract}

\maketitle

%%%%%%%%%%%%%%%%%%%%%%%%%%%%%%%%%%%%%%%%%%%%%%%%%%%%%%%%%%%%%%%%%%%%%%%%%%%%%%%%
%
% Introduction
%
%%%%%%%%%%%%%%%%%%%%%%%%%%%%%%%%%%%%%%%%%%%%%%%%%%%%%%%%%%%%%%%%%%%%%%%%%%%%%%%%

\section{Introduction}

%%%%%%%%%%%%%%%%%%%%%%%%%%%%%%%%%%%%%%%%%%%%%%%%%%%%%%%
% Thermal rectification
%%%%%%%%%%%%%%%%%%%%%%%%%%%%%%%%%%%%%%%%%%%%%%%%%%%%%%%%

Different thermal rectification mechanisms for heat radiation in the nanoscale regime have been studied which of course also, in general, work in far-field setups. One mechanism is to use the intrinsic temperature dependence of the material properties of the involved materials~\cite{FanRectification2010, Iizuka2012, BasuEtAl2011, WangEtAl2013} which can lead to a very large heat flux rectification, i.e.\ a diode-like effect, when the involved elements undergo a phase transition as superconductors~\cite{Nefzaoui2014, OrdonezEtAl2017} or phase change materials like VO$_2$ as studied theoretically and experimentally in Refs.~\cite{PBASAB2013,Yangetal2013,ItoEtAl,FiorinoEtAl2018}. Such diode like elements can also function as thermal transistors for heat radiation~\cite{PBASABtransistor} as experimentally verified in a far-field setup~\cite{Yungui}. Interestingly, also asymmetric many-body systems can show a thermal rectification~\cite{Ivan}.

%%%%%%%%%%%%%%%%%%%%%%%%%%%%%%%%%%%%%%%%%%%%%%%%%%%%%%%
%Nonreciprocal
%%%%%%%%%%%%%%%%%%%%%%%%%%%%%%%%%%%%%%%%%%%%%%%%%%%%%%%%
Another possibility to achieve thermal rectification for heat radiation is to employ magneto-opcial materials which have a non-reciprocal material response when an external magnetic field is applied. This non-reciprocal property results in interesting properties of near- and far-field thermal radiation. For example it could be shown that in thermal equilibrium there is a persistent heat current~\cite{Zhu2016}, spin and angular momentum~\cite{Silveirinha,Ottcircular2018}, a Hall effect for heat radiation~\cite{PBAHall2016}, spin-directional thermal emission~\cite{DongEtAl2021}, a giant magneto-resistance~\cite{Latella2017,Cuevas}, a Berry phase of thermal radiation~\cite{SABPBABerry2022} as well as a non-reciprocal version of the Green-Kubo relation~\cite{Herz}. Also the impact on the heat radiation between planar films and multilayers with an applied magnetic field has been studied~\cite{MoncadaEtAl2015,FanEtAl2020,QuEtAl2024} as well as the heat flux between a magneto-optical nanoparticle and planar substrate~\cite{Raul2}. Finally, the coupling to non-reciprocal surface modes allows for a strong heat flux rectification~\cite{Ottdiode2019} due to the spin-spin coupling~\cite{OttSpin2020} to the non-reciprocal surface waves which are spin-momentum locked~\cite{Mechelen,Zubin2019}. All those works show that the magnetic field provides a mean to control the magnitude and direction of the radiative heat flux actively(see also reviews~\cite{OttReview2019,SABRMP2021}). 

%%%%%%%%%%%%%%%%%%%%%%%%%%%%%%%%%%%%%%%%%%%%%%%%%%%%%%%%
%Weyl
%%%%%%%%%%%%%%%%%%%%%%%%%%%%%%%%%%%%%%%%%%%%%%%%%%%%%%%%%%
The existence of intrinsically non-reciprocal materials such as the Weyl semimetals (WSM)~\cite{ChenEtAl2016,Kotov2018} allows for realizing the same effects as found for magneto-optical materials but without the need of an externally applied magnetic field. Consequently, it could be shown that there is a photonic spin hall effect~\cite{DaEtAl2021}, an anomalous Hall effect for thermal radiation~\cite{OttAnomalous2020}, modulated heat transfer between slabs and multilayers~\cite{TangEtAl2021,XuEtal2020,YuEtAlMultilayer2022,YuEtAl3B2022,Sheng2023}, negative differential thermal conductance~\cite{SunEtAl}, the possibility to enable thermal routing~\cite{GuoEtAl2020}, non-reciprocal absorption and emission~\cite{ZhaoEtAl2020,ZhangZhu2023}, broadband circularly polarized thermal radiation~\cite{Zubin}, non-reciprocal far-field heat radiation~\cite{WuEtAl2022}, and coupling to graphene nanoribbons~\cite{YuEtAl2023}. These and other effects are reviewed in Ref.~\cite{GuoEtAl2023,DidariEtAl2024}, for instance. Of course, there is also a thermal rectification due to the coupling to non-reciprocal surface~\cite{HuEtAl2023}. The heat flux rectification is predicted to be more efficient for WSM than for magneto-optical materials like InSb~\cite{HuEtAl2023}.

\begin{figure}
	\centering
	\includegraphics[width = 0.45\textwidth]{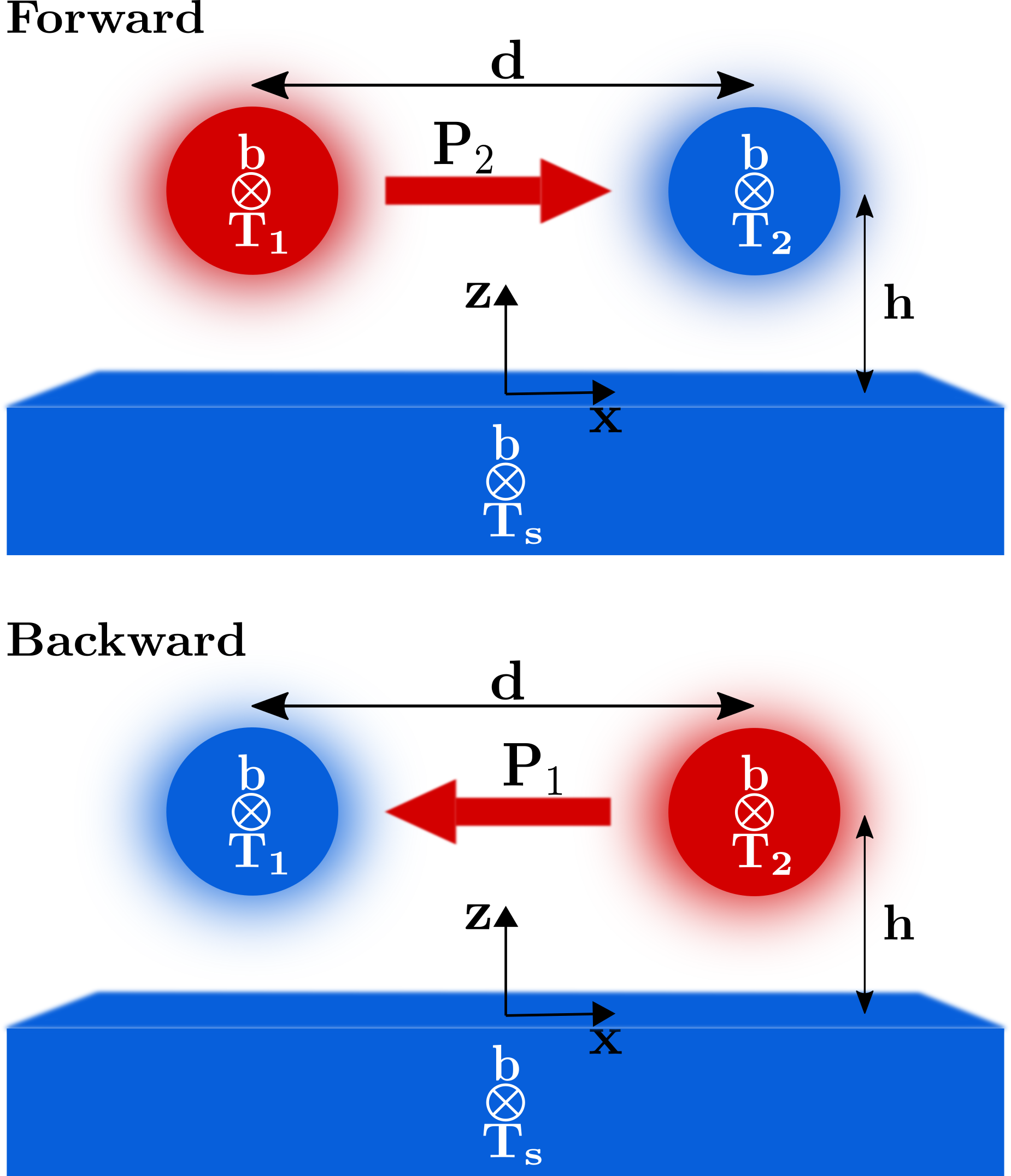}\\
	\caption{Sketch of our physical system consisting of two spherical nanoparticles with Radius $R$ and temperatures $T_1$, $T_2$ placed at $-d/2$ and $d/2$ in a height $h$ above a planar substrate. The substrate and the nanoparticles are made of the same WSM and the temperature of the substrate is at temperature $T_s$. In the forward case $T_1 > T_2$ and $T_s = T_2$ whereas in the backward case $T_2 > T_1$ and $T_s = T_1$. The Weyl node separation $\mathbf{b} = b \hat{y}$ is aligned in positive $y$ direction.}
	\label{Fig:Sketch}
\end{figure}

The aim of our work is to provide a detailed study of the thermal rectification of the heat flux between two WSM nanoparticles by the interaction with the non-reciprocal surface waves of a close by planar WSM sample as sketched in Fig.~\ref{Fig:Sketch} and first studied for InSb in Refs.~\cite{Ottdiode2019,OttSpin2020} and for WSM in Ref.~\cite{HuEtAl2023}. We explain the underlying mechanism on the basis of the properties of the involved surface modes and show that the thermal rectification ratios found in Ref.~\cite{HuEtAl2023} and in particular the value of 2673 are incorrect. In the same range of parameters as in  Ref.~\cite{HuEtAl2023} we find much lower values than in the original work~\cite{HuEtAl2023}. However, by changing the interparticle distance we show that rectification ratios of about 6000 can be obtained. Furthermore, we show that a regime of high rectification for small values of the Weyl node separation has been overlooked previously, which can easily lead to rectification ratios of about 8000. Our results are substantiated with an analysis of the contribution of the surface modes which qualitatively explains our observations. Finally, in a parameter analysis we demonstrate that even rectification ratios of 15000 are realizable in the studied parameter regime underlining, on the one hand, the fact that WSMs are in general strong candidates for heat flux rectification without the necessity of applying external magnetic fields and on the other our analysis helps to identify the best WSM for a strong heat flux rectification.

Our work is structured in the following way: In Sec.~II we introduce the heat flux expression for two non-reciprocal nanoparticles above a non-reciprocal planar sample within dipole approximation. In Sec.~III we define the used permittivity of WSMs and introduce the parameters used for numerical calculations. The localized resonances in the nanoparticles and the surface waves in the planar sample are discussed in Sec.~IV followed by a detailed discussion of the heat flux rectification in Sec.~V. Finally, in Sec.~VI study the impact of the Weyl nodes separation, the number of Weyl nodes, Fermi energy, dissipation, and temperature on the rectification ratio.

%%%%%%%%%%%%%%%%%%%%%%%%%%%%%%%%%%%%%%%%%%%%%%%%%%%%%%%%%%%%%%%%%%%%%%%%%%%%%%%%
%
% NFRHT Weyl-Semi-Metals
%
%%%%%%%%%%%%%%%%%%%%%%%%%%%%%%%%%%%%%%%%%%%%%%%%%%%%%%%%%%%%%%%%%%%%%%%%%%%%%%%%

\section{Near-field radiative heat transfer}

Within the framework of fluctuational electrodynamics the radiative heat flux
between dipolar objects each at local thermal equilibrium has been studied intensively
and general expressions for the mean power received by each dipolar object have been 
derived~\cite{SABRMP2021}. Considering only two dipolar objects having local equilibrium 
temperatures $T_1$ and $T_2 < T_1$ in close vicinity to an interface at 
temperature $T_s = T_2$ (see forward case in Fig.~\ref{Fig:Sketch}) the power $P_2$ received by object $2$ 
has the general form~\cite{SABRMP2021}
\begin{equation}
   P_2 = 3 \int_0^\infty \!\! \frac{\rd \omega}{2\pi} \hbar \omega n_1 \mathcal{T}_{12}
\end{equation}
with the mean photon number $n_1 = (\re^{\hbar\omega/\kb T_1} - 1)^{-1}$, a transmission coefficient $\mathcal{T}_{12}$, and Planck's reduced constant $\hbar$. In the reverse scenario where $T_2 > T_1$ and $T_s = T_1$ (see backward case in Fig.~\ref{Fig:Sketch}) the power $P_1$ received by object $1$ has the general form
\begin{equation}
   P_1 = 3 \int_0^\infty \!\! \frac{\rd \omega}{2\pi} \hbar \omega n_2 \mathcal{T}_{21}.
\end{equation}
with mean photon number $n_2 = (\re^{\hbar\omega/\kb T_2} - 1)^{-1}$ and a transmission coefficient $\mathcal{T}_{21}$. In the single scattering limit the transmission coefficients are given by simple trace formulas ($i,j = 1,2$)~\cite{SABRMP2021}
\begin{equation}
	\mathcal{T}_{ij} = \frac{4}{3} k_0^4 \Tr\bigl[ \boldsymbol{\chi}_j \mathds{G}(\mathbf{r}_j, \mathbf{r}_i) \boldsymbol{\chi}_i \mathds{G}^\dagger(\mathbf{r}_j, \mathbf{r}_i)  \bigr].
	\label{Eq:TransmissionCoefficient}
\end{equation}
In these expressions for both transmission coefficients the Green function $\mathds{G}(\mathbf{r}_j, \mathbf{r}_i)$ contains the whole information about the environment. In our case the Green function consists of the vacuum and a scattering part due to the presence of a substrate. For a planar non-reciprocal semi-infinite substrate the explicit expressions for the Green function can be found in Ref.~\cite{Ottdiode2019,HuEtAl2023}, for instance. Now, the properties of the two dipolar objects enter into the transmission coefficient by 
\begin{equation}
	\boldsymbol{\chi}_i \approx \frac{1}{2 \ri} \bigl( \boldsymbol{\alpha}_i - \boldsymbol{\alpha}_i^\dagger \bigr)
\end{equation}
where $\boldsymbol{\alpha}_i$ is the polarizability tensor of the dipolar object. Note, that here we neglect the radiation correction which is typically small in the infrared regime. Obviously, the non-hermitian part of the polarizability of the two objects determines the heat transfer between the objects. For spherical nanoparticles with radii $R_1 = R_2 \equiv R$ as discussed here, the polarizabilities are given in terms of the particles permittivity tensors $\boldsymbol{\epsilon}_i$ ($i = 1,2$) as~\cite{SABRMP2021}
\begin{equation}
	\boldsymbol{\alpha}_i =  4 \pi R^3(\boldsymbol{\epsilon}_i - \mathds{1}) (\boldsymbol{\epsilon}_i + 2\mathds{1})^{-1}
\end{equation}
In the following we will use the above approximate expressions for the numerical evaluation of the radiative heat transfer in order to have comparable results to Ref.~\cite{HuEtAl2023} where the same approximations have been used. The goal is to study the heat transfer between the nanoparticles via the surface waves of a nearby interface~\cite{Saaskilathi2014,Asheichyk2017,DongEtAl2018,paper_2sic,ZhangEtAl2019,HeEtAl2019,LiuEtAl2023} but with the specificity that the surface waves are non-reciprocal~\cite{Ottdiode2019,OttSpin2020}. We finally emphasize that the dipolar approximation only holds when the distance between the dipolar objects and the dipolar objects and the interface are sufficiently large. As a rule of thumb a center-to-center (center-to-surface distance) of $4R$ ($3R$) is sufficient to guarantee the validity of the dipolar approximation~\cite{Naraynaswamy2008,Otey,Becerril}. Therefore, in our numerical calculations we restrict ourselves to a surface to particle distance of 5R and interparticle distances larger than 3R.

%%%%%%%%%%%%%%%%%%%%%%%%%%%%%%%%%%%%%%%%%%%%%%%%%%%%%%%%%%%%%%%%%%%%%%%%%%%%%%%%
%
% Permittivity and Surface Waves in Weyl-Semi-Metals 
%
%%%%%%%%%%%%%%%%%%%%%%%%%%%%%%%%%%%%%%%%%%%%%%%%%%%%%%%%%%%%%%%%%%%%%%%%%%%%%%%%

\section{Permittivity of WSM}

Our main focus is on the radiative heat transfer between WSM nanoparticles due to the coupling to 
non-reciprocal surface modes in a close by substrate of the same WSM. For these materials the permittivity
tensor is non-reciprocal, i.e.~$\boldsymbol{\epsilon} \neq \boldsymbol{\epsilon}^t$. Assuming that the 
vector connecting the Weyl nodes in k-space is pointing in y-direction and the WSM possess no chiral magnetic effect, the permittivity tensor is 
\begin{equation}
	\boldsymbol{\epsilon}(T,\omega) = \begin{pmatrix} \epsilon_d & 0 & \ri \epsilon_a \\ 0 & \epsilon_d & 0 \\ -\ri \epsilon_a & 0 & \epsilon_d\end{pmatrix}.
\end{equation}

\begin{figure}
	\centering
	\includegraphics[width = 0.45\textwidth]{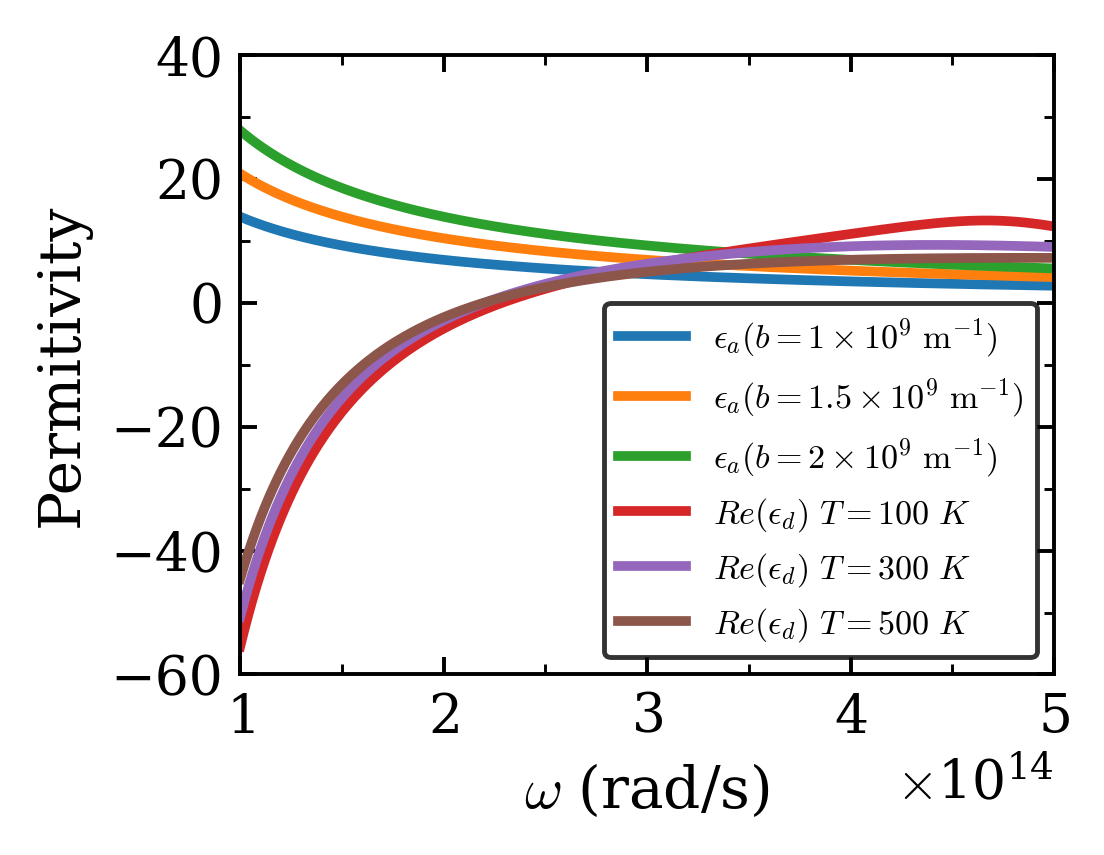}\\
	\caption{The permittivity tensor components $\epsilon_a $ and $\epsilon_d$ of the WSM as function of frequency. For the non-diagonal elements $b$ is varied whereas for the diagonal components we vary the temperature.}
	\label{Fig:eps_weyl}
\end{figure}

In a generally accepted simplified model the diagonal and non-diagonal elements are~\cite{Kotov2018}
\begin{equation}
	\epsilon_a(\omega) = \frac{b e^2}{2\pi^2 \epsilon_0 \hbar \omega}
\end{equation}
and
\begin{equation}
\begin{split}
	\epsilon_d (\omega,T) &= \epsilon_b + \ri \frac{r_s g_w \Omega}{6 \Omega_0} G\biggl(\frac{\Omega}{2}\biggr) \\
	&\quad- \frac{r_s g_w}{6 \pi \Omega_0} \biggl\{ \frac{4}{\Omega} \biggl[ 1 + \frac{\pi^2}{3} \biggl( \frac{\kb T}{E_F} \biggr)^2 \biggr] \\ &\qquad + 8 \Omega \int_0^{\eta_c}\!\!\! \rd \eta\, \eta \frac{G(\eta) - G(\Omega/2)}{\Omega^2 - 4 \eta^2} \biggr\}.
\end{split}
	\label{Eq:EpsilonWeyl}
\end{equation}
Here we have introduced the general quantities like the permittivity of vacuum $\epsilon_0$ and the model and material dependent quantities $\Omega = \hbar(\omega + \ri \tau^{-1})/E_F$, $\Omega_0 = \hbar \omega/E_F$, $r_s = e^2/(4 \pi \epsilon_0 \hbar v_F)$, the Fermi level $E_F$, the Fermi velocity $v_F$, the number of Weyl nodes $g_w$, the separation of the Weyl nodes $2b$, the cutoff energy $E_c$ and $\eta_c = E_c/E_F$. Finally, we also have $G(E) = n(-E) - n(E)$ where $n(E)$ is the Fermi distribution function. As material parameters we use $g_w = 2$, $\epsilon_b = 6.2$, $v_F = 0.83\times10^5 {\rm m/s}$, $\tau = 10^{-12}\,{\rm s}$, $b = 2\times10^9 {\rm m}^{-1}$, $\eta_c = 3$ and the temperature dependent Fermi level
\begin{widetext}
\begin{equation}
   	E_F(T)=\frac{2^{1/3}\bigg[9E_F(0)^3+\sqrt{81E_F(0)^6+12\pi^6k_B^6T^6}\bigg]^{2/3}-2\pi^2 3^{1/3}k_B^2T^2}{6^{2/3}\bigg[9E_F(0)^3+\sqrt{81E_F(0)^6+12\pi^6k_B^6T^6}\bigg]^{1/3}}
\end{equation}
\end{widetext}
with $E_F(0{\rm K}) = 0.163\,{\rm eV}$ so that $E_F(300{\rm K}) = 0.15\,{\rm eV}$. This choice of parameters is the same as in Ref.~\cite{HuEtAl2023} and corresponds to typical doped WSM like Co$_3$MnGa~\cite{YuEtAlMultilayer2022}. A similar choice of parameters but with $b = 8.5\times10^{8}\,{\rm m}^{-1}$ corresponds to Co$_3$Sn$_2$S$_2$~\cite{Sheng2023} (see also table~\ref{table:TabulatedWSMParams}). Interestingly, in some WSMs the Weyl node separation $2b$ can also be tuned electrically, magnetically, and optically~\cite{TsaiEtAl2020,RayEtAl2020,LiEtAL2020}.

In Fig.~\ref{Fig:eps_weyl} we show the diagonal and non-diagonal components of the permittivity for different temperatures and different values of the Weyl node separation $b$.  We therewith reproduce the same results as in Ref.~\cite{HuEtAl2023}. It can be observed that the non-diagonal elements are very sensitive to a change in $b$ whereas the diagonal elements depend on temperature only.

%%%%%%%%%%%%%%%%%%%%%%%%%%%%%%%%%%%%%%%%%%%%%%%%%%%%%%%%%%%%%%%%%%%%%%%%%%%%%%%%
%
% Surface Waves in Weyl-Semi-Metals 
%
%%%%%%%%%%%%%%%%%%%%%%%%%%%%%%%%%%%%%%%%%%%%%%%%%%%%%%%%%%%%%%%%%%%%%%%%%%%%%%%%

\section{Localized surface modes and non-reciprocal surface waves in Weyl-Semi-Metals}

The heat or energy transfer between the particles is due to a coupling of the localized resonances in the particle with the surface mode resonances
of the planar substrate. The localized resonances of the nanoparticles are determined by the poles of the polarizability and here we obtain as for 
magneto-optical materials with applied magnetic field three dipolar resonances with magnetic quantum numbers $m = 0, \pm 1$ defined by
\begin{equation}
	\epsilon_d(\omega_{m = 0}) = -2
\end{equation}
and
\begin{equation}
	\epsilon_d(\omega_{m = \pm 1}) = - 2 \mp  \epsilon_a(\omega_{m = \pm 1}). 
	\label{Eq:LocalizedResonances}
\end{equation}
Obviously, the non-diagol elements induce a Zeeman-like splitting~\cite{Ottcircular2018,Raul1} linear in $\epsilon_a$ and therefore linear in the Weyl node separation $b$ whereas the resonance frequency $\omega_{m = 0}$ does not depend on $b$. In table~\ref{table:ParticleResonance} we list some numerical results for the resonance frequencies for different temperatures and Weyl node separations. Due to the obvious analogy of the analytical expressions, one can find for the WSM non-vanishing spin, angular momentum and a circular heat flux for the resonance $\omega_{m = \pm 1}$ as discussed for magneto-optical materials in Ref.~\cite{OttReview2019,Ottcircular2018}. Due to the temperature dependence of the diagonal elements of the permittivity tensor of the WSM the exact spectral position of the resonance frequencies will depend not only on $b$ but also on the temperature of the nanoparticles.

\begin{table}
\begin{tabular}{|c|c|c|c|c|}
	\hline
	$T$ & $b $ & $\omega_{m = 0}$ & $\omega_{m = +1}$ & $\omega_{m = -1}$ \\ 
	 (K) & $ (10^{9}\,{\rm m}^{-1})$ & $(10^{14}\,{\rm rad/s})$ & $(10^{14}\,{\rm rad/s})$ & $(10^{14}\,{\rm rad/s})$ \\ \hline \hline
	 200 & 2 & 2.08 & 1.36 & 3.08 \\ \hline
	 180 & 2 & 2.09 & 1.37 & 3.08 \\\hline
	 180 & 0.5 & 2.09 & 1.88 & 2.3 \\\hline
	 180 & 0.19 & 2.091 & 2.008 & 2.176 \\\hline
\end{tabular}
	\caption{ \label{table:ParticleResonance} Numerical values of the nanoparticle resonance frequencies $\omega_{m = 0, \pm 1}$ for particle temperatures of $180\,{\rm K}$ and $200\,{\rm K}$ and different values of the Weyl node separation.}
\end{table}

\begin{figure*}[hbt]
	\centering
	\includegraphics[width = 0.9\textwidth]{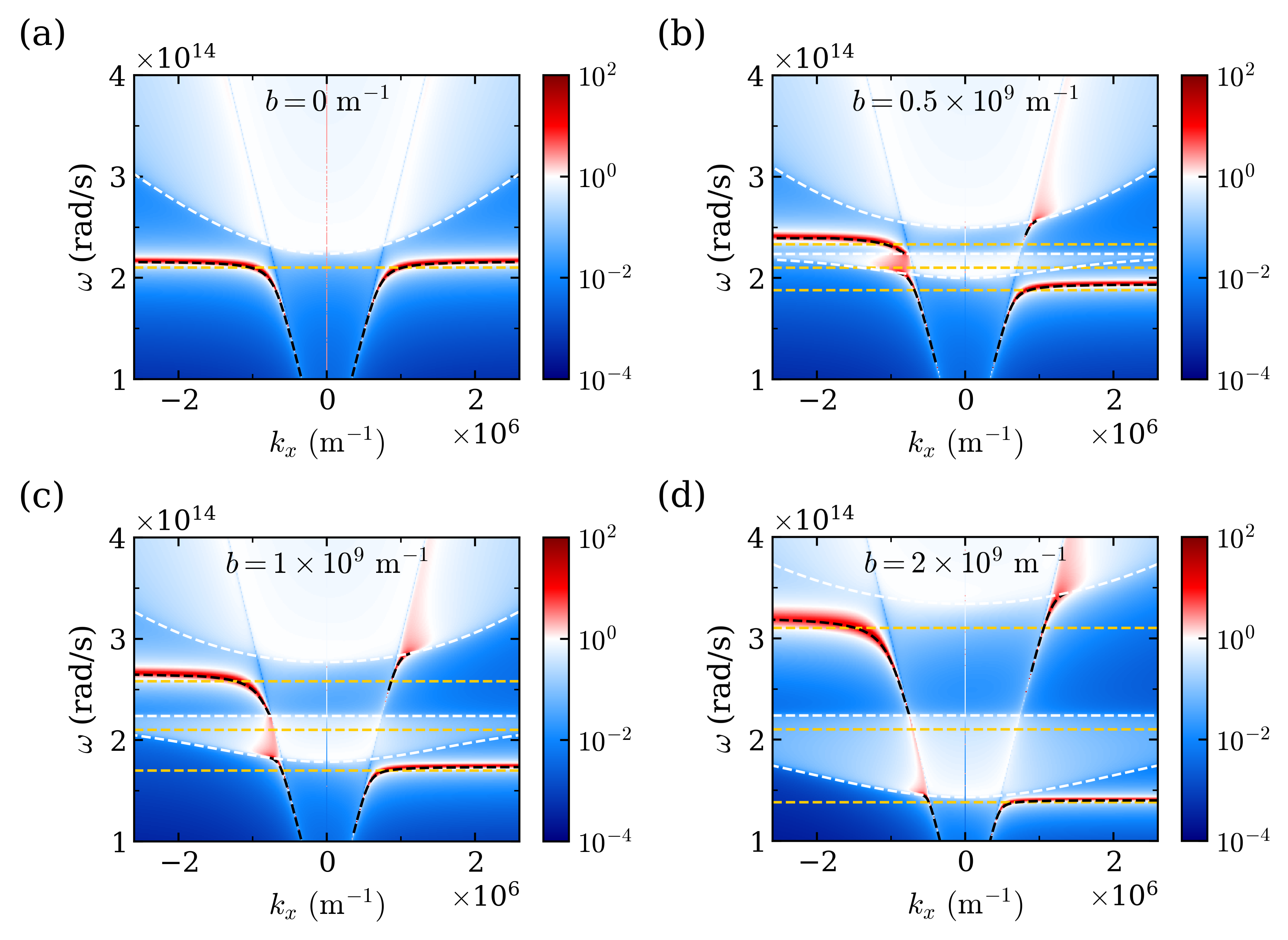}
	 \caption{Reflection coefficient for p-polarized light $r_{pp}$ of Weyl material for the Voigt configuration in $\omega$-$k_x$ plane for different values of $b = 0, 0.5\times 10^9, 1\times 10^9, 2\times 10^9{m^{-1}}$ for $T_s = 180\,{\rm K}$. For propagating waves with $|k_x| \leq k_0 $ the quantitiy $1-|r_{pp}|^2$ is shown and for evanescent waves with $|k_x| > k_0$ the quantity $\Im(r_{pp})$ is plotted. The black dashed lines are the dispersion relations of the surface modes from Eq.~(\ref{Eq:Dispersionrelation}). The white lines are the light lines defined by $k_x = k_0 \sqrt{\epsilon_V}$. The horizontal dashed orange lines mark the frequencies of the resonances $\omega_{m = 0, \pm 1}$ of the nanoparticles determined from Eq.~(\ref{Eq:LocalizedResonances}).}
	\label{Fig:rpp_weyl}
\end{figure*}

On the other hand, the interface of the planar WSM will support surface waves which are non-reciprocal and have similar properties as the non-reciprocal surface waves in magneto-optical materials with applied magnetic field~\cite{Hofmann,Kotov2018}. For these non-reciprocal surface waves general expressions for the dispersion relation can be derived~\cite{BrionEtAl1972,wallis,chiuquinn}. Here we are mainly interested in the propagation of the surface waves in $\pm x$ direction assuming a Voigt configuration, i.e.\ an applied magnetic field or Weyl node vector perpendicular to the $x$-$z$ plane spanned by the nanoparticles and the surface. In that case ($k_y = 0$) the surface wave dispersion relation can be simplified to~\cite{chiuquinn}
\begin{equation}
	k_x^2 - k_0^2 \epsilon_d - k_z k_z^V \epsilon_d - k_z k_x \ri \epsilon_a = 0
	\label{Eq:Dispersionrelation}
\end{equation}
with the wavevectors $k_z = \sqrt{k_0^2 - k_x^2}$, $k_z^V = \sqrt{k_0^2 \epsilon_V - k_x^2}$  and the Voigt permittivity $\epsilon_V = (\epsilon_d^2 - \epsilon_a^2)/\epsilon_d$. To determine the corresponding $k_x$ and $\omega$ values of the dispersion relation of surface modes we determine real valued solutions $k_x$ for real frequencies. This can be done when neglecting dissipation in the material. On the other hand, to determine the propagation length we determine the complex valued solutions of $k_x$ for real frequencies. In this case, the dissipation inside the material is included. Then we can define the surface propagation length in a standard way by~\cite{OttSpin2020}
\begin{equation}
	\Lambda_{\pm} = \frac{1}{2 \Im(k_x^{\pm})}
	\label{Eq:Proplength}
\end{equation}
where $k_x^{\pm}$ is the complex solution of the dispersion relation for waves travelling in $\pm x$ direction.

In Fig.~\ref{Fig:rpp_weyl} we show $1 - |r_{pp}|^2$ for the propagating waves with $|k_x| < k_0$ and $\Im(r_{pp})$ the evanescent waves with $|k_x| > k_0$ using the expressions for the reflection coefficient for anisotropic media from Ref.~\cite{chen}. The plotted quantities in the $\omega-k_x$ plane are essentially proportional to the photonic density of states. The different surface wave modes in the evanescent sector $|k_x| > k_0$ with large values for $\Im(r_{pp})$ can be nicely seen as well as the Zeeman splitting of surface waves travelling towards $\pm x$ direction for non-zero Weyl node separation $b \neq 0$. For $b = 0$ there is no non-reciprocity and the three dipolar resonances as well as the surface mode resonances are degenerate. This effect of the splitting of the surface waves is the same as for magneto-optical surface waves when a magnetic field is applied~\cite{Ottdiode2019,Hofmann,Kotov2018,BrionEtAl1972,wallis,chiuquinn} so that the Weyl node separation $\mathbf{b} = b \hat{y}$ plays the same role as the magnetic field $\mathbf{B} = B \hat{y}$. It should be noted that the dispersion relations of the surface modes shown in Fig.~\ref{Fig:rpp_weyl} are determined by neglecting losses in order to avoid backbending effects, whereas the reflection coefficients are evaluated with taking dissipation into account. By comparing the dispersion relations obtained from Eq.~(\ref{Eq:Dispersionrelation}) with the reflection coefficients it can be seen that with dissipation the surface modes can exist at frequencies which are forbidden in the case without dissipation, i.e.\ in the region between the white dashed Voigt lightlines defined by $k_x = k_0 \sqrt{\epsilon_V}$. In particular, there are surface modes travelling in negative $x$ direction in the frequency band between the two Zeeman splitted surface modes. Furthermore, we have plotted in Fig.~\ref{Fig:rpp_weyl} the frequencies of the localized dipolar resonances within the nanoparticles. From this representation it can be seen which particle resonances can couple to the surface waves in $\pm x$ direction. This feature will be essential to understand the rectification mechanism for the radiative heat transfer. 

\begin{figure}
	\centering
	\includegraphics[width = 0.45\textwidth]{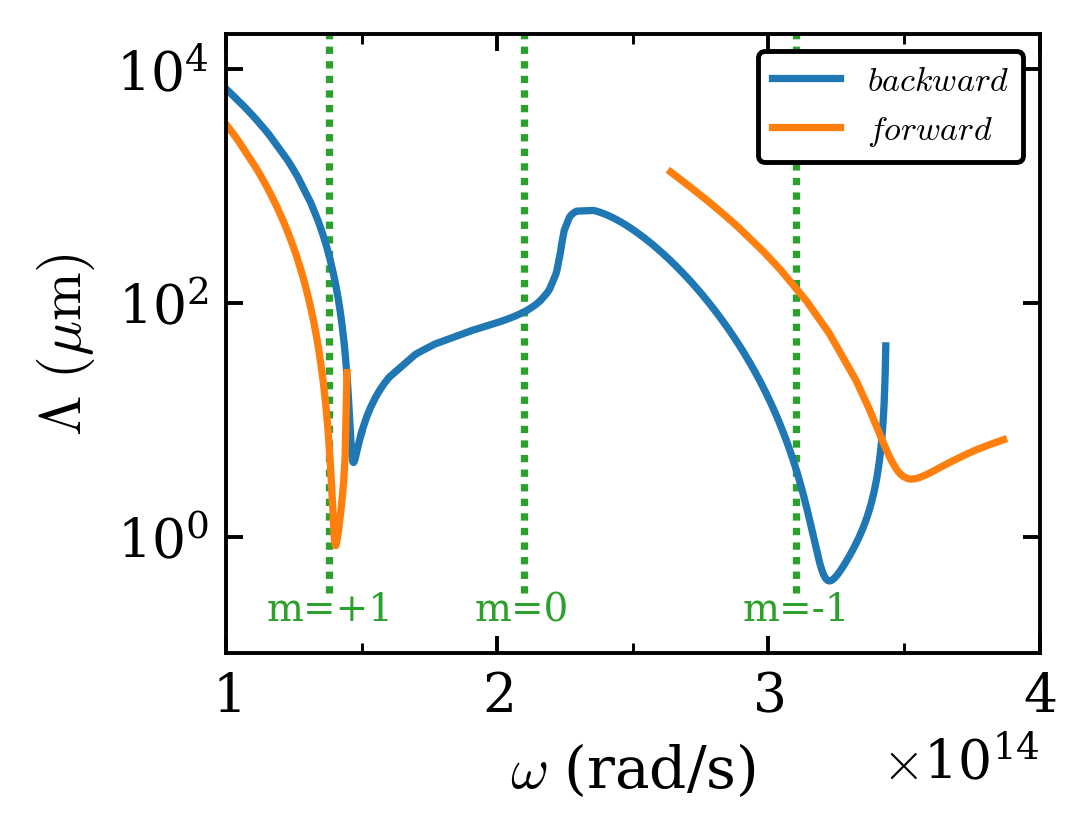}
	\caption{Propagation length $\Lambda_\pm$ from Eq.~(\ref{Eq:Proplength}) for the Weyl point separation $b = 2\times10^{9}\,{\rm m}^{-1}$. Vertical lines mark the positions of the localized particle resonances.
	}
	\label{Fig:Proplength}
\end{figure}

In Fig.~\ref{Fig:Proplength} we show the propagation length $\Lambda_{\pm}$ of the surface waves traveling in $\pm x$ obtained from Eq.~(\ref{Eq:Proplength}) for $b = 2\times10^{9}\,{\rm m}^{-1}$ with losses included. As can already be seen in Fig.~\ref{Fig:rpp_weyl}(d) the localized resonance at $\omega_{m = 0}$ can only couple to surface waves travelling in $-x$ direction, because there are no surface wave solutions for surface waves travelling in $+x$ direction for this frequency. On the other hand, the resonances $\omega_{m = \pm 1}$ can couple to surface waves travelling in both $\pm x$ direction but with the difference that the propagation length is dependent on the propagation direction. At $\omega_{m = + 1}$ the excited surface waves in $-x$ direction travel farther than those in $+x$ direction and at  $\omega_{m = -1}$ it is the opposite trend. From this observation one can expect that for large enough interparticle distance $d$ the backward heat flow is preferred at  $\omega_{m = + 1}$ and a forward heat flow at $\omega_{m = -1}$.

%%%%%%%%%%%%%%%%%%%%%%%%%%%%%%%%%%%%%%%%%%%%%%%%%%%%%%%%%%%%%%%%%%%%%%%%%%%%%%%%
%
% Heat Flux Rectification
%
%%%%%%%%%%%%%%%%%%%%%%%%%%%%%%%%%%%%%%%%%%%%%%%%%%%%%%%%%%%%%%%%%%%%%%%%%%%%%%%%

\section{Heat Flux rectification}

The heat flux rectification between the nanoparticles can be quantified by the rectification ratio~\cite{HuEtAl2023}
\begin{equation}
	\eta = \frac{P_1 - P_2}{P_2}
	\label{Eq:eta}
\end{equation}
which by definition can be larger than one. In some works like Ref.~\cite{PBASAB2013} normalized rectification ratios are used. This definition is useful when the heat flow is enhanced in backward direction. In the opposite case when the heat flux is enhanced in forward direction then it is better to use the rectification ratio
\begin{equation}
	\tilde{\eta} = \frac{P_2 - P_1}{P_1}
	\label{Eq:etatilde}
\end{equation}
correspondingly.

\begin{figure}
	\centering
	\includegraphics[width = 0.45\textwidth]{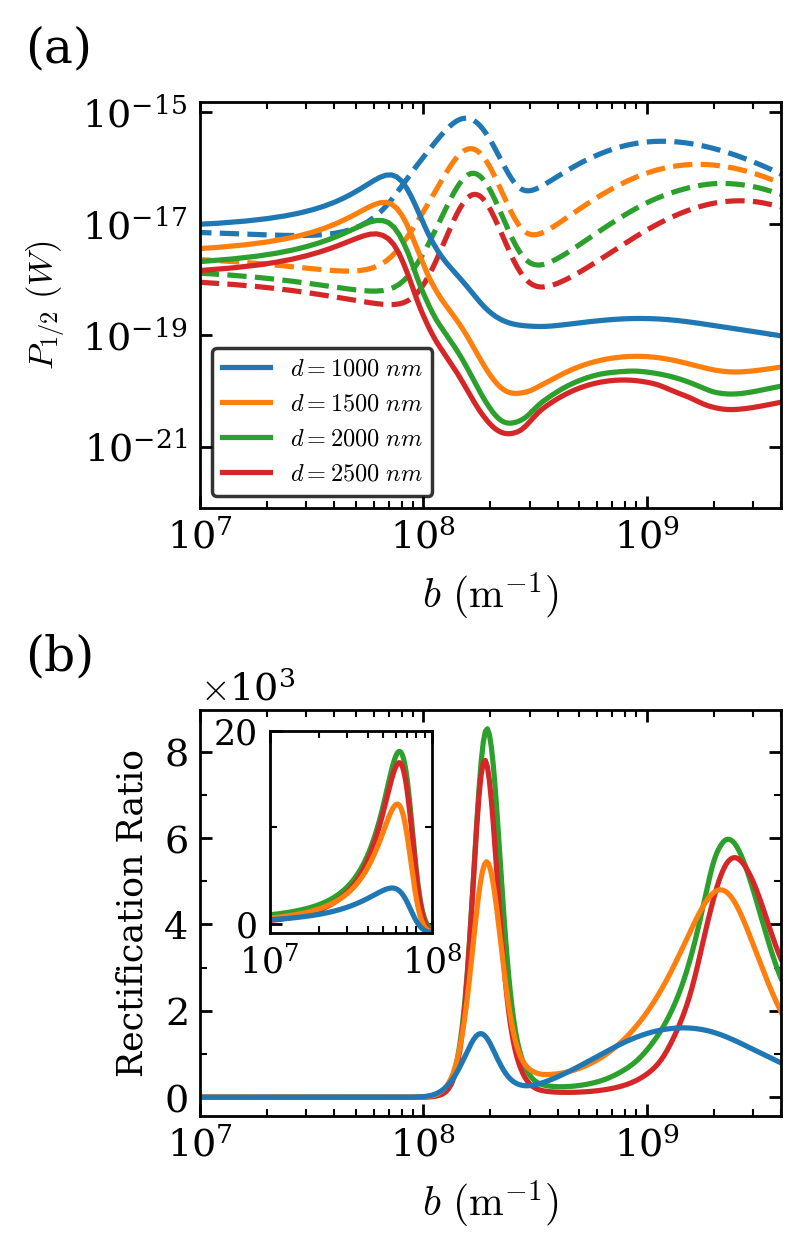}
	\caption{(a) $P_2$ (solid line) and $P_1$ (dashed line) and (b) Rectification ratio $\eta$ from Eq.~(\ref{Eq:eta}) between two WSM nanoparticles of Radius $R = 20\,{\rm nm}$ above a WSM substrate at a height of $h = 100\,{\rm nm}$ as a function of the Weyl node separation for different interparticle distance $d = 1000\,{\rm nm}, 1500\,{\rm nm}$, $2000\,{\rm nm}$, and $2500\,{\rm nm}$. Inset: Rectification ratio $\tilde{\eta}$ from Eq.~(\ref{Eq:etatilde}) for the forward rectification.}
	\label{Fig:eta_weyl_distance}
\end{figure}

In Fig.~\ref{Fig:eta_weyl_distance} we show the numerical results for $P_1$ and $P_2$ in the backward and forward scenario together with the rectification ratio $\eta$ for two WSM nanoparticles of Radius $R = 20\,{\rm nm}$ at a distance of $h = 100\,{\rm nm}$ above a WSM substrate as a function of the Weyl node separation $b$. We chose $T_1=200\,{\rm K}$ , $T_2=T_s=180\,{\rm K}$ for the forward and vice versa for the backward case. It can be seen that there are for the shown distances in general two maxima for the power $P_1$ and $P_2$ as well as for the rectification ratio. The location of the maxima depends on the distance $d$. We note that for values of $b$ smaller than $10^{8}\,{\rm m}^{-1}$ the power $P_2$ for the forward case is larger than $P_1$ for the backward case. This means that a heat flux in forward direction is preferred and therefore this rectification can hardly be seen in the rectification ratio $\eta$ but in $\tilde{\eta}$ which can have values up to 15 for the studied distances [see inset of Fig.~\ref{Fig:eta_weyl_distance}(b)]. This rectification in forward direction is small compared to the strong rectification in backward direction for larger values of $b$ as can be nicely seen in Fig.~\ref{Fig:eta_weyl_distance}(b). The first rectifcation maximum for small $b$ can be as large as about 8000 and the second rectification maximum at larger values of $b$ can be as large as about 6000 for $d = 2000\,{\rm nm}$. 

\begin{figure}
	\centering
	\includegraphics[width = 0.45\textwidth]{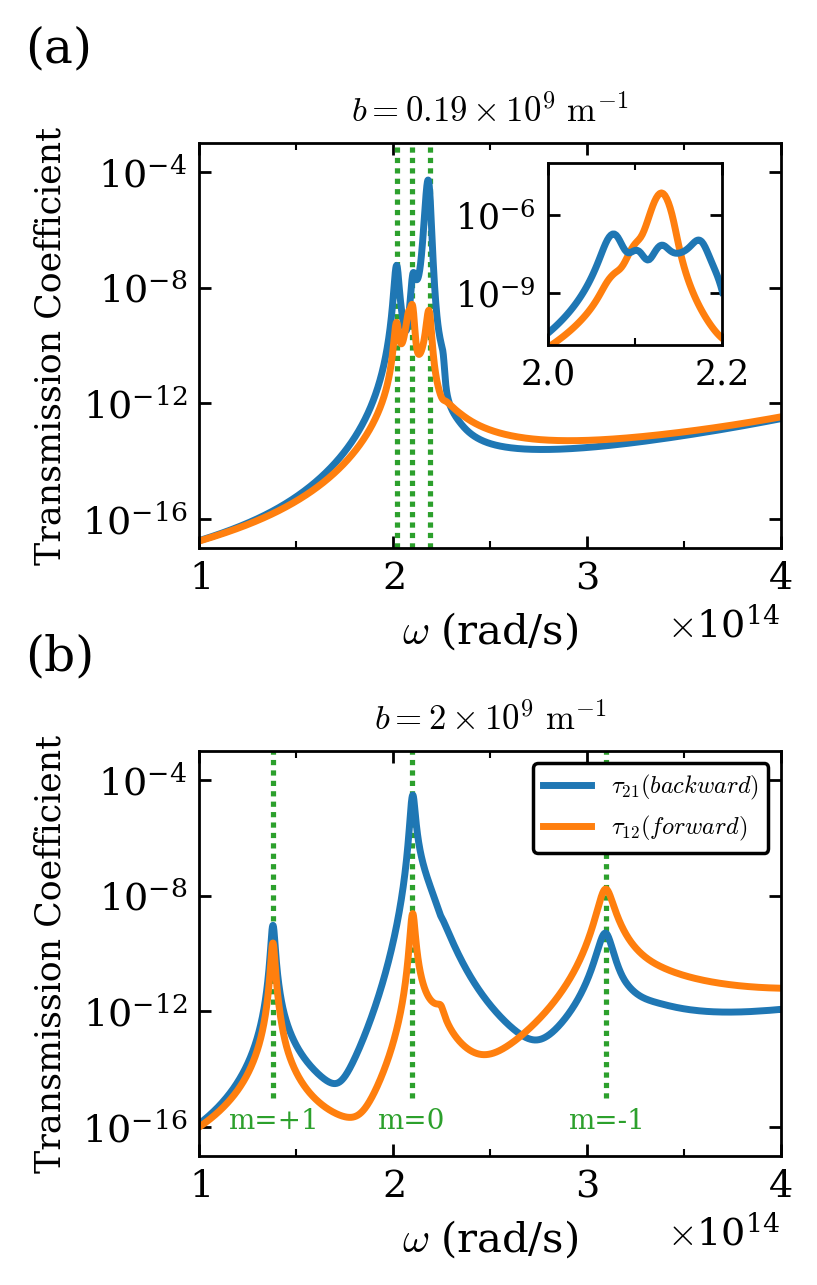}
	\caption{Transmission coefficients $\mathcal{T}_{12}$ and $\mathcal{T}_{21}$ for the backward and forward case for $d = 1500\,{\rm nm}$ for (a) $b = 0.19\times10^{9}\,{\rm m}^{-1}$ and $b = 2\times10^{9}\,{\rm m}^{-1}$. The dashed lines indicate the spectral positions of the particle resonances. Inset: $\mathcal{T}_{12}$ and $\mathcal{T}_{21}$ for $b = 6\times10^{7}\,{\rm m}^{-1}$. }
	\label{Fig:Transmission}
\end{figure}

In order to understand the physical mechanism leading to the large rectification in backward direction for two distinct values of $b$ we study in more detail the transmission coefficients $\mathcal{T}_{12}$ and $\mathcal{T}_{21}$ introduced for the forward and backward case in Eq.~(\ref{Eq:TransmissionCoefficient}). In Fig.~\ref{Fig:Transmission} we show both transmission coefficients chosing the interparticle separation distance $d = 1500\,{\rm nm}$ for $b = 0.19\times10^{9}\,{\rm m}^{-1}$ and $b = 2\times10^{9}\,{\rm m}^{-1}$, i.e.\ for those values of the Weyl node separation where we find a maximal rectification ratio in Fig.~\ref{Fig:eta_weyl_distance}. It can be nicely seen that there are three peaks due to the localized resonances within the nanoparticles corresponding the the resonance frequencies $\omega_{m = 0, \pm 1}$ for the given value of $b$. For $b = 0.19\times10^{9}\,{\rm m}^{-1}$ the transmission cofficient is in the backward direction larger than in the forward direction at all three resonances. On the other hand, for $b = 2\times10^{9}\,{\rm m}^{-1}$ the two low frequency resonances are enhanced in the backward direction, whereas the high frequency resonance is enhanced for the forward direction. 

\begin{figure}
	\centering
	\includegraphics[width = 0.45\textwidth]{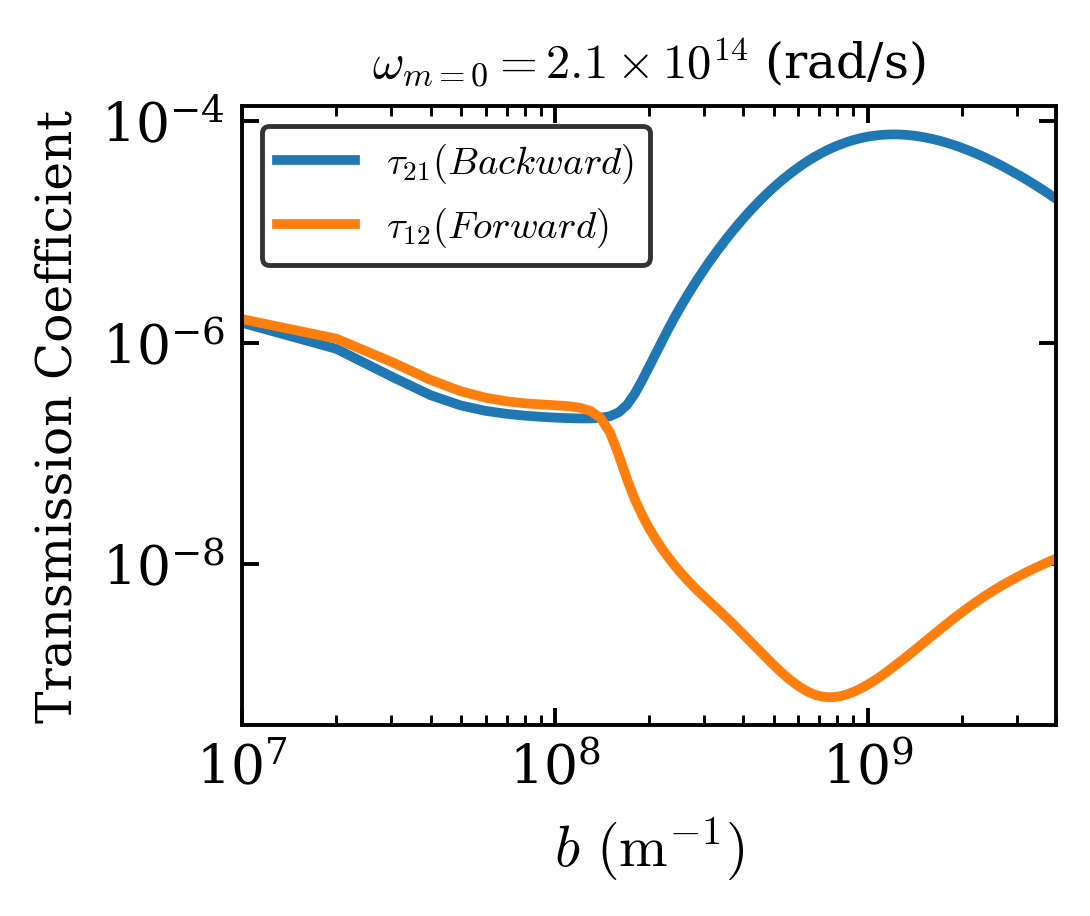}
	\caption{Transmission coefficient $\mathcal{T}_{12}$ and $\mathcal{T}_{21}$ for the backward and forward case for $d = 1000\,{\rm nm}$ evaluated at $\omega_{m = 0}$ as function of the Weyl node separation $b$.}
	\label{Fig:TransmissionM0}
\end{figure}

Since the rectification is due to the heat exchange between the nanoparticles via the coupling to the surface modes, it can be explained by the properties of the surface modes. From the reflection coefficient in Fig.~\ref{Fig:rpp_weyl}(a) and (b) it becomes obvious that for relatively small values of $b$ the Zeeman-like splitting is relatively small. Actually, for very small $b < 10^8\,{\rm m}^{-1}$ the particle resonances at $\omega_{m = 0}$ and $\omega_{m = -1}$ can couple to the surface modes in forward, i.e.\ $+x$ direction. This leads to a predominant rectification in forward direction as can be seen in the curves of $P_2$ in Fig.~\ref{Fig:eta_weyl_distance}(a) and the transmission coefficients in the inset of Fig.~\ref{Fig:Transmission}(a). For $b > 10^8\,{\rm m}^{-1}$ a gap opens between the upper and lower bands of surface mode solutions as demonstrated in the reflection coefficient in Fig.~\ref{Fig:rpp_weyl}(b). Due to this gap opening, the two high frequency particle resonances and in particular the resonance at $\omega_{m = 0}$ cannot anymore couple to surface modes in forward direction so that $P_2$ decreases for $b > 10^8\,{\rm m}^{-1}$. Note, that for much larger values of $b$ the resonance at $\omega_{m = -1}$ can again couple to surface modes so that mainly the behaviour of the resonance at $\omega_{m = 0}$ decides about the dominance of the heat flux in forward and backward direction. For $\omega_{m = 0}$ we show in Fig.~\ref{Fig:TransmissionM0} the transmission coefficient as function of $b$ clearly demonstrating the transition at about $b = 0.19\times10^{9}\,{\rm m}^{-1}$ from heat transfer in forward direction to heat transfer in backward direction. At this transition point all particle resonances can efficiently couple to surface modes in backward direction as can be seen in the transmission coefficient in Fig.~\ref{Fig:Transmission} (a) whereas the coupling of the three particle resonances to the surface modes in forward direction is generally weaker.

	For larger values of $b = 2\times10^{9}\,{\rm m}^{-1}$ as shown in Fig.~\ref{Fig:Transmission} (b) the dominance for the heat flow in backward direction indeed stems from the resonance at $\omega_{m = 0}$ and the contribution of the other resonances is still strong in both direction as these resonances can couple to surface modes in both directions as seen in the reflection coefficient in Fig.~\ref{Fig:rpp_weyl}(d) and the transmission coefficient in Fig.~\ref{Fig:Transmission} (b). From the propagation length in Fig.~\ref{Fig:Proplength}(b) we found that the propagation length is higher in backward direction for the resonance at $\omega_{m = +1}$ and higher in forward direction for the resonance at $\omega_{m = -1}$. This feature explains the larger transmission coefficient at $\omega_{m = +1}$ in backward direction and the larger transmission coefficient in forward direction at  $\omega_{m = -1}$. We emphasize that of course also the coupling strength of the resonances plays an important role as studied in more detail in Ref.~\cite{OttSpin2020}.
%\textcolor{red}{It can also be understood by the spin-spin coupling mechanism observed in Ref.~\cite{OttSpin2020}: The high frequency particle resonance at $\omega_{m = -1}$ preferably couples to the surface waves in forward direction and the low frequency particle resonance at  $\omega_{m = +1}$ couples preferably in backward direction. This behavior can be nicely seen in Fig.~\ref{Fig:Transmission}(b). The resonance at $\omega_{m = 0}$ has no preference but here due to the fact that there are no surface waves at this frequency in forward direction the heat flux in backward direction is preferred.} The fact, that for $b =0.19\times10^{9}\,{\rm m}^{-1}$ all nanoparticle resonances are enhanced in backward direction but for $b = 2\times10^{9}\,{\rm m}^{-1}$ only the two low frequency resonances explains why the rectification maximum is higher for lower values of $b$.  

For very large $b \gg 10^{9}\,{\rm m}^{-1}$ the resonances $\omega_{m = \pm 1}$ move out of the spectral window determined by the Planck function so that in this case one might expect that there remains the rectification due to the coupling of the resonance at $\omega_{m = 0}$ to the surface modes in backward direction. Therefore, the rectification tends to zero without changing its sign for large values of $b$ in contrast to what was found for magneto-optical materials for large magnetic fields where the rectification changes its sign for large magnetic fields. Of course, one can also argue that we have a similar effect of changing the direction of rectification at relatively low $b$ around $10^{-8}\,{\rm m}^{-1}$.

\begin{figure}
	\centering
	\includegraphics[width = 0.45\textwidth]{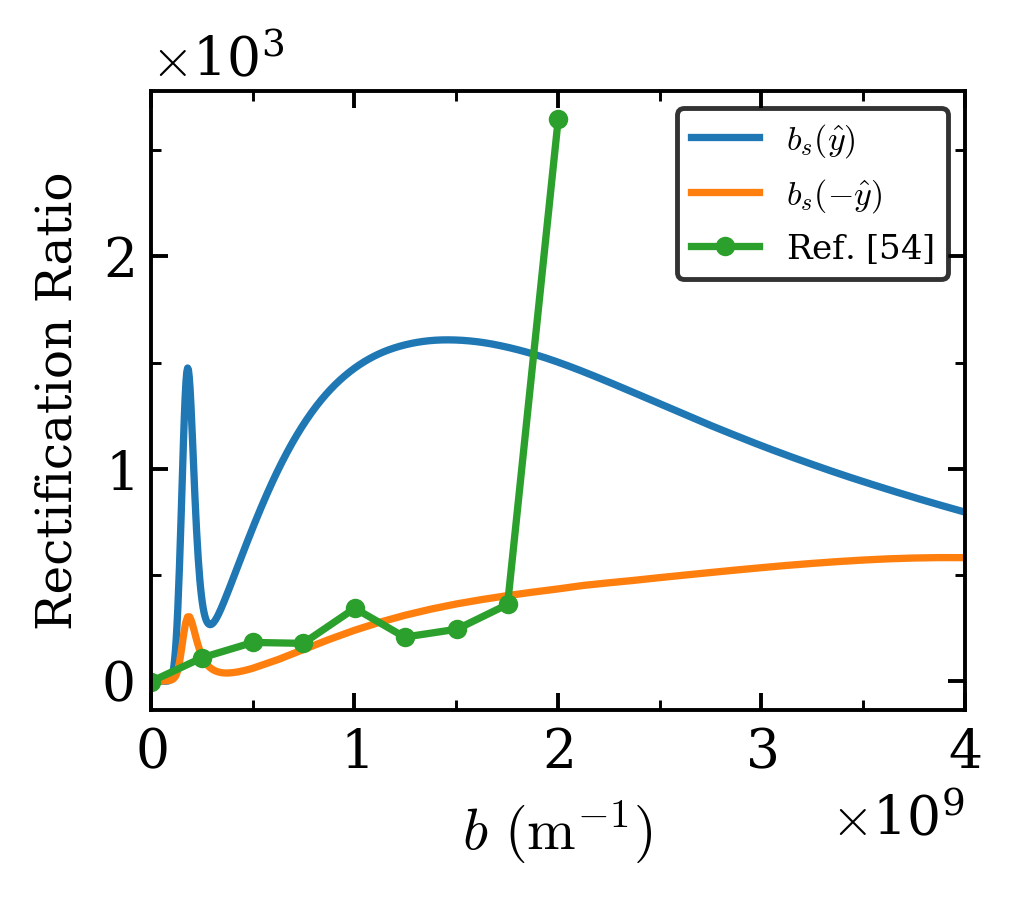}
	\caption{Rectification ratio $\eta$ from Eq.~(\ref{Eq:eta}) for $\mathbf{b} = b \hat{y}$ (blue line) in the substrate in positive y direction and $\tilde{\eta}$ from Eq.~(\ref{Eq:etatilde}) for $\mathbf{b} = - b \hat{y}$ (orange line) in the substrate in negative y direction as a function of the Weyl node separation for the interparticle distance $d = 1000\,{\rm nm}$. In the particles we have in both cases $\mathbf{b} = b \hat{y}$. For comparision we the values of Ref.~\cite{HuEtAl2023} (green line) which correspond to the configuration for $\mathbf{b} = b \hat{y}$ (blue line).}
	\label{Fig:eta_weyl_distanceb}
\end{figure}

Finally, one can of course change the direction of rectification by turing the sample by $180^\circ$ such that the orientation of the weyl nodes is in the particles in $+y$ direction and in the sample in $-y$ direction. However, the coupling strength between the surface modes and the particle resonances is reduces in this case so that the rectication changes its sign, but the rectification ratio is reduced as well as can be seen in Fig.~\ref{Fig:eta_weyl_distanceb}. The general features seen in the rectification ratio remain the same. For comparision we have also added the numerical valus of Ref.~\cite{HuEtAl2023} which correspond to the case that all Weyl nodes are oriented in  $+y$ direction (blue curve). It can be seen that these numerical values are different from our values. In our opinion, the large scattering in the numerical values of Ref.~\cite{HuEtAl2023} indicate some numerical problems as well as the sudden jump for $b = 2\times10^{9}\,{\rm m}^{-1}$ to a rectification ratio of $2673$. We find rather a value of 1502 for $b = 2\times10^9\,{\rm m}^{-1}$. Therefore, even though qualitatively it is true that WSM are good heat flux rectifiers the main numerical findings in Ref.~\cite{HuEtAl2023} are incorrect. Additionally, in contrast to Ref.~\cite{HuEtAl2023} we do not find a more or less monotonic increase of the rectification ratio with $b$ which could as in Ref.~\cite{HuEtAl2023} be explained by the permittivity ratio which linearly increases with $b$. Indeed, the mechanism is much more complex as we have detailed above. Furthermore, due to the fact that only a few data points are evaluated in Ref.~\cite{HuEtAl2023} and that the focus was only on $\eta$ and not $\tilde{\eta}$ the two high rectification regimes for small Weyl nodes separations have been overlooked.

\begin{figure}
	\centering
	\includegraphics[width = 0.45\textwidth]{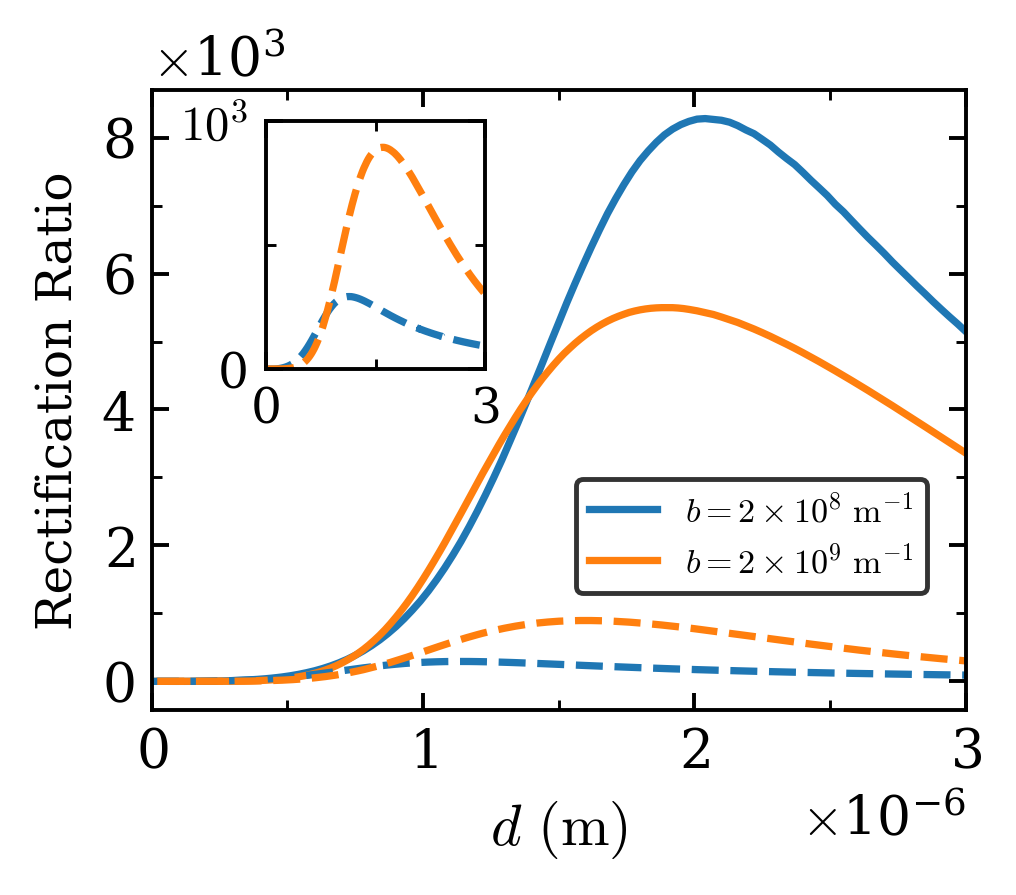}
	\caption{Rectification ratio $\eta$ (solid line) from Eq.~(\ref{Eq:eta}) as function of distance $d$ for or $b = 2\times 10^{8}\,{\rm m}^{-1}$ and or $b = 2\times 10^{9}\,{\rm m}^{-1}$ as well as $\tilde{\eta}$ (dashed line) from Eq.~(\ref{Eq:etatilde}) for $\mathbf{b} = - b \hat{y}$ in the substrate, i.e.\ the substrate is rotated by 180$^\circ$, for $b = 2\times 10^{8}\,{\rm m}^{-1}$ and $b = 2\times 10^{9}\,{\rm m}^{-1}$ . Inset: Shows a blow up of the rectification ratios $\tilde{\eta}$. }
	\label{Fig:eta_diffd}
\end{figure}

In Fig.~\ref{Fig:eta_diffd} we show the variation of the rectification ratios as function of distance for $b = 2\times 10^{8}\,{\rm m}^{-1}$ and $b = 2\times 10^{9}\,{\rm m}^{-1}$ (for both orientations of $\mathbf{b} = \pm b \hat{y}$ in the sample). From these curves one can read off the maximum possible values of rectification in 4 cases: (a) $\eta = 8282$ for $b = 2\times 10^{8}\,{\rm m}^{-1}$ at a distance of $2.04\,\mu{\rm m}$ and (b) $\eta = 5498$ for $b = 2\times 10^{9}\,{\rm m}^{-1}$  at a distance of $1.9\,\mu{\rm m}$ both when the orientation of the Weyl node separation is $\mathbf{b} = b \hat{y}$, and (c) $\tilde{\eta} = 292$ for $b = 2\times 10^{8}\,{\rm m}^{-1}$ at a distance of $1.14\,\mu{\rm m}$, and (d) $\tilde{\eta} = 895$ for $b = 2\times 10^{9}\,{\rm m}^{-1}$  at a distance of $1.6\,\mu{\rm m}$ both when the orientation of the Weyl node separation in the substrate is $\mathbf{b} = - b \hat{y}$. Hence, by turning the sample by $180^\circ$ it is possible to inverse the direction of the preferred heat flow with still substantial rectification ratios.

%%%%%%%%%%%%%%%%%%%%%%%%%%%%%%%%%%%%%%%%%%%%%%%%%%%%%%%%%%%%%%%%%%%%%%%%%%%%%%%%
%
% Parameter analysis 
%
%%%%%%%%%%%%%%%%%%%%%%%%%%%%%%%%%%%%%%%%%%%%%%%%%%%%%%%%%%%%%%%%%%%%%%%%%%%%%%%%

\section{Parameter analysis}

The material parameters of a selection of different WSMs taken from Ref.~\cite{Sheng2023} are tabulated
in table~\ref{table:TabulatedWSMParams}. It can be seen that these four materials have already a certain spread in the different parameters and most notably in the Fermi energy at zero temperature $E_F(0\,{\rm K})$ and in the number of Weyl points $g$. In the following we will study how the rectification ratio changes as function of $E_F(0\,{\rm K})$ for different Weyl node separations and number of Weyl points. From the expression of the permittivity in Eq.~(\ref{Eq:EpsilonWeyl}) it can be expected that the impact of the value of $E_F(0\,{\rm K})$ and $g$ are rather high, because they determine the spectral position of the zeros in the real part of the permittivity and therefore the spectral position of the surface modes and localized modes which are essential for a strong heat flux rectification. In this section we use the parameters for the typical WSM studied in the previous sections and starting from that we change only a few parameters as explicitely mentioned and calculate the rectification ratio. This study of the sensitivity of the rectification ratio on the material parameters can help to select the corresponding WSM for obtaining the highest rectification ratio for a certain distance and temperature regime.

\begin{table}
\begin{tabular}{|c|c|c|c|c|c|}
	\hline
	WSM & $b (10^9 {\rm m}^{-1})$ & $\tau$ (fs) & $E_F(0\,{\rm K})$ (eV) & $g$ & $v_{\rm F} (10^{5} {\rm m/s})$ \\ \hline
	Co$_2$MnGa & $2.12$ & 450 & 0.06 & 12 & 0.83 \\
	Co$_3$Se$_2$S$_2$ & 2 & 600 & 0.11 & 12 & 2 \\
	Co$_3$Sn$_2$S$_2$ & 0.85 & 1000 & 0.163 & 2 & 0.83 \\
	Eu$_2$IrO$_7$     & 2.7  & 500  & 0.01 & 24 & 10 \\ \hline
        typical WSM & 2 & 1000 & 0.163 & 2 & 0.83 \\ \hline
\end{tabular}
	\caption{ \label{table:TabulatedWSMParams} Material parameters for different WSM taken from Ref.~\cite{Sheng2023} and the Material studied here and in Ref.~\cite{HuEtAl2023} which corresponds typical doped WSM like Co$_3$MnGa~\cite{YuEtAlMultilayer2022}.}
\end{table}

\begin{figure}
	\centering
	\includegraphics[width = 0.45\textwidth]{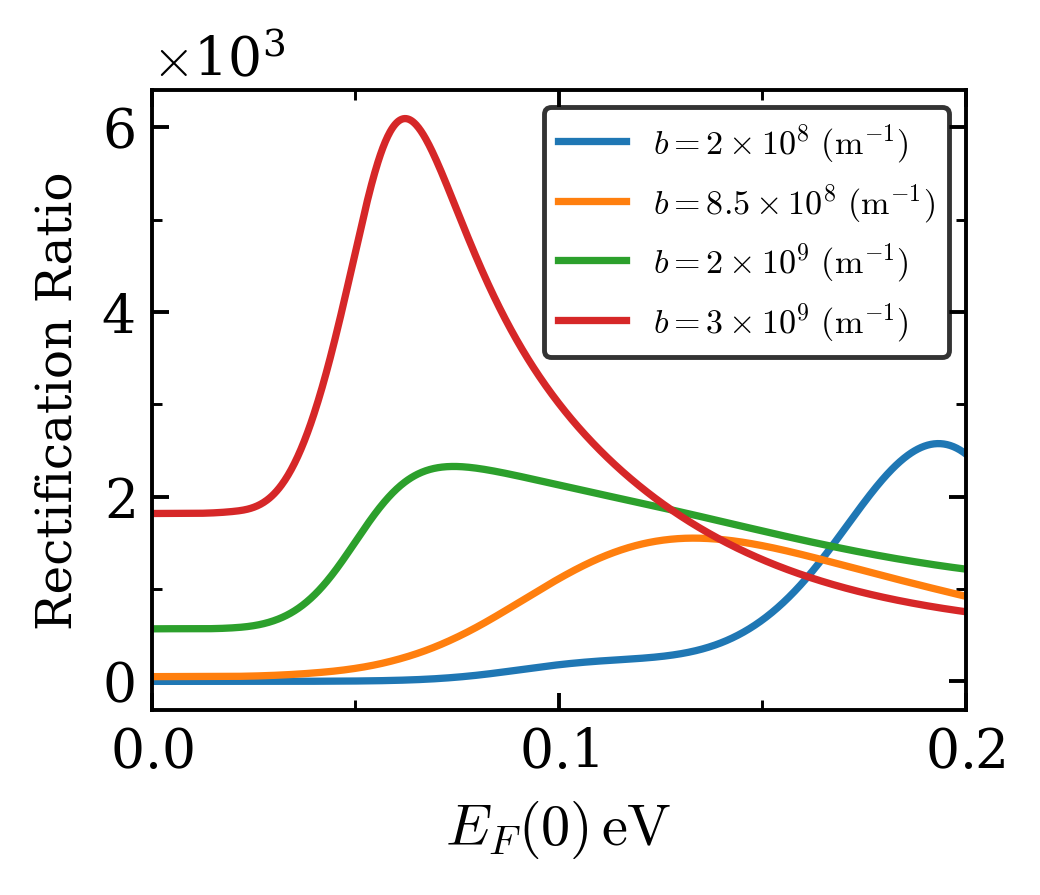}
	\caption{Rectification ratio $\eta$ from Eq.~(\ref{Eq:eta}) as function of the Fermi energy $E_F(0\,{\rm K})$ for an interparticle distance of $d = 1000$nm  and four different values of the Weyl nodes separation $b$ which correspond to values of the WSMs tabulated in table~\ref{table:TabulatedWSMParams}.
	\label{Fig:Fermi}}
\end{figure}

	First, we study the impact of the zero temperature Fermy energy $E_F(0\,{\rm K})$ in Fig.~\ref{Fig:Fermi} for the different values of the Weyl node separation $b$ such that we cover a large enough parameter range which includes the values of $E_F(0\,{\rm K})$ and $b$ for the different WSM tabulated in table~\ref{table:TabulatedWSMParams}.
	From the numerical results it can be seen that for the chosen interparticle distance the rectification ratio is very sensitive to the Fermi energy as well as the Weyl node separation. There seems to be a maximum for all curves which shifts towards larger values of the Fermi energy when the Weyl node separation $b$ is decreased. Furthermore, the numerical values indicate that theoretically even much larger rectification ratios than the maximum of about 1600 found in Fig.~\ref{Fig:eta_weyl_distanceb} for $E_F(0 \,{\rm K}) = 0.163{\rm eV}$ are possible for WSM with a smaller value of the Fermi energy.

\begin{figure}
	\centering
	\includegraphics[width = 0.45\textwidth]{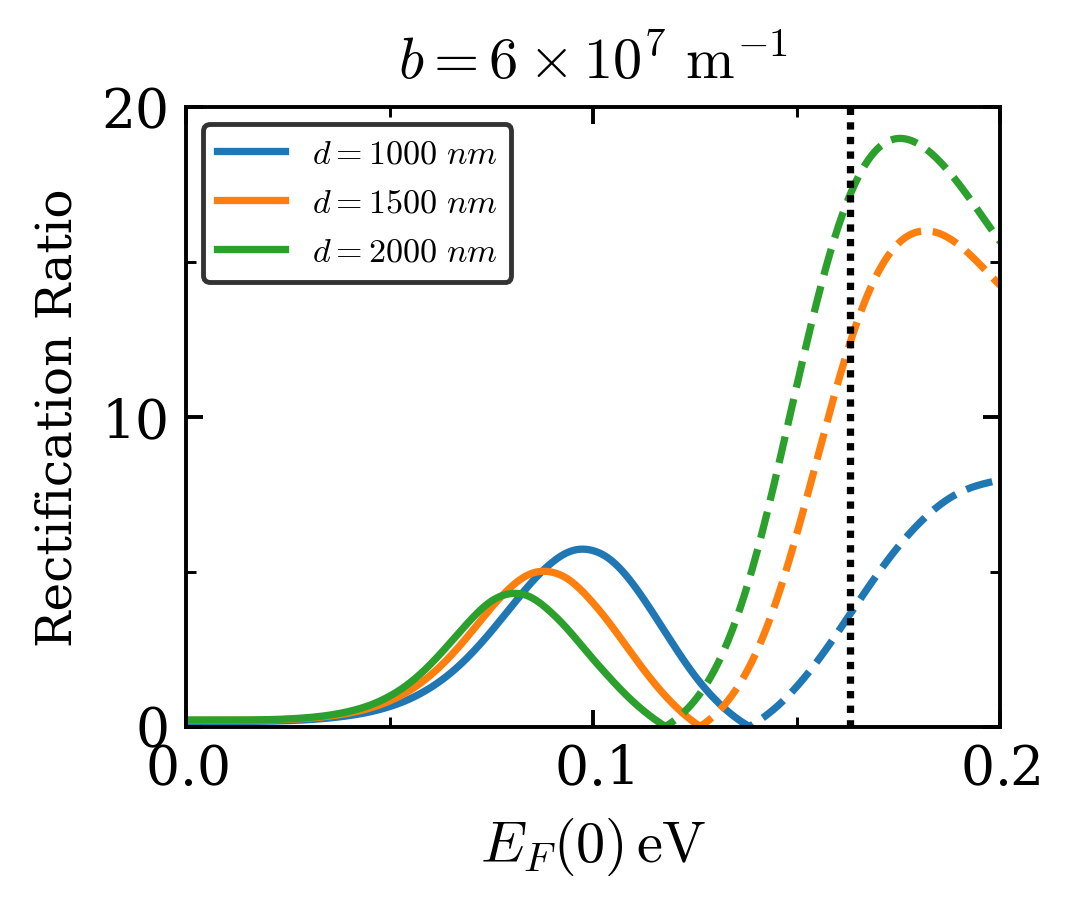}
	\caption{Rectification ratio $\eta$ (solid lines) from Eq.~(\ref{Eq:eta}) and $\tilde{\eta}$ (dashed lines) from Eq.~(\ref{Eq:etatilde}) as function of the Fermi energy $E_F(0\,{\rm K})$ for a interparticle distance of $d = 1000$nm, 1500nm, 2000nm and $b = 6\times10^{7}\,{\rm m}^{-1}$. The vertical line marks the value of $E_F(0\,{\rm K}) = 0.163\,{\rm eV}$ as used in the previous sections.
	\label{Fig:Fermi6e9}}
\end{figure}

\begin{figure}
	\centering
	\includegraphics[width = 0.45\textwidth]{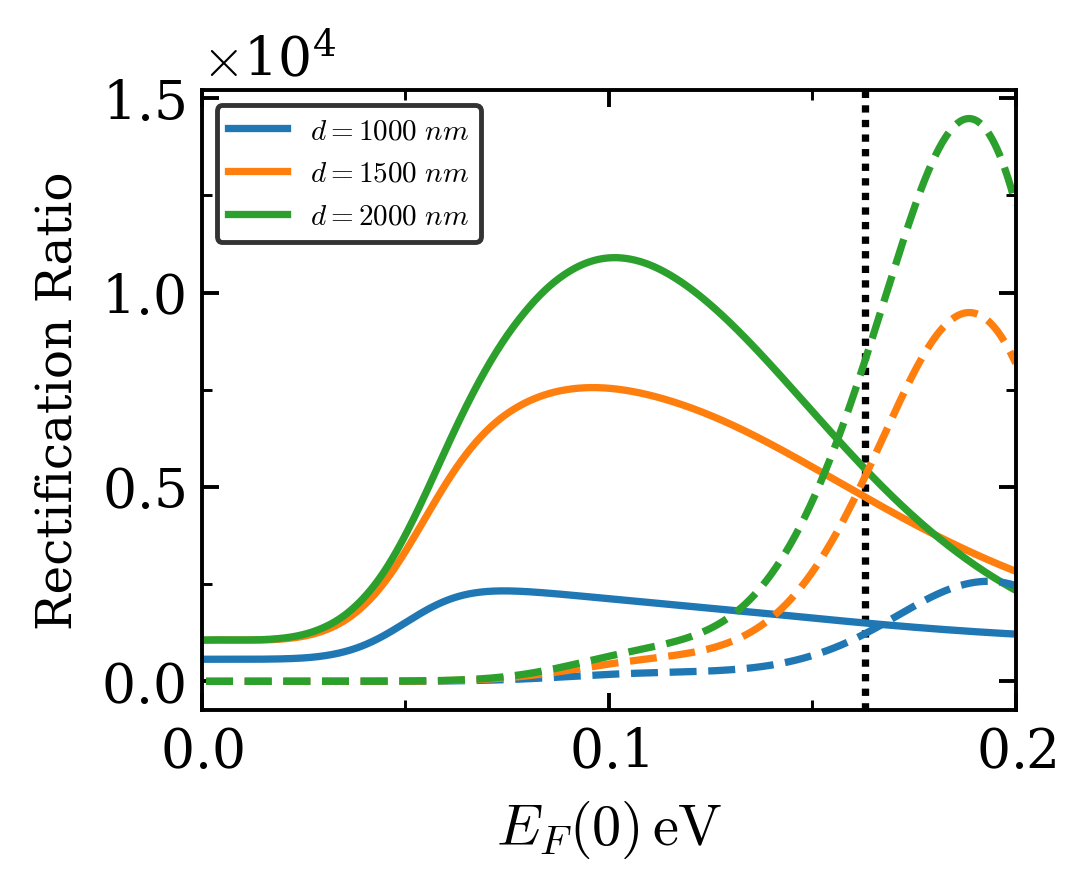}
	\caption{Rectification ratio $\eta$ from Eq.~(\ref{Eq:eta}) as function of the Fermi energy $E_F(0\,{\rm K})$ for an interparticle distance of $d = 1000$nm, 1500nm, 2000nm and two Weyl node separations $b = 2\times10^{8}\,{\rm m}^{-1}$ (dashed lines) and $b = 2\times10^{9}\,{\rm m}^{-1}$ (solid lines). The vertical line marks the value of $E_F(0\,{\rm K}) = 0.163\,{\rm eV}$ as used in the previous sections.
	\label{Fig:Fermi2e9}}
\end{figure}

	We further study the sensitivity of the rectification ratio $\eta$ for backward rectification and $\tilde{\eta}$ for forward rectification for small values of Weyl node separations in Fig.~\ref{Fig:Fermi6e9}. It can be seen that for $E_F(0\,{\rm K}) = 0.163\,{\rm eV}$ indeed there is a forward rectification of the heat flux as discussed in the previous sections. On the other hand, one can observe that for smaller values of the Fermi energy $E_F(0\,{\rm K})$ one can also find a predominant backward rectification. However, in both cases the rectification is relatively small for the chosen distance and temperature. In Fig.~\ref{Fig:Fermi2e9} we further show the rectification ratio $\eta$ for larger values of $b = 2\times10^8\,{\rm m}^{-1}$ and $b = 2\times10^9\,{\rm m}^{-1}$ where we find a dominant rectification in backward direction. Again a strong sensitivity to the Fermi energy $E_F(0\,{\rm K})$ can be observed and very large rectification ratios up to 15000 are found. This again demonstrates that WSM are strong candidates for a highly efficient heat flux rectification.

Let us have a closer look at some candidates. Ferromagnetic WSM possess their magnetic property for temperatures smaller than the Curie temperature which is for Co$_3$Se$_2$S$_2$, Co$_3$Sn$_2$S$_2$ about 150K-177K~\cite{Kotov2018} which is close to the temperature regime studied in Ref.~\cite{HuEtAl2023} and this work. However, there are also magnetic room temperature WSMs like Co$_2$MnGa with a Curie temperature of 690K~\cite{RoomtemperatureWSM,COMNGA}. Another good candidate for another room temperature WSM is the compensated ferrimagnet Ti$_2$MnAl which has a very high Curie temperature of 650K and similar parameters as the antiferromagnetic Eu$_2$IrO$_7$~\cite{Kotov2018} which has a magnetic order below its N\'{e}el temperature of 120-150K. Therefore the temperature range for which WSMs are non-reciprocal is relatively large and consequently it is useful to study the rectification ratio as a function of temperature as well. To this end, we set now for the forward case $T_1 = T_s + \Delta T$ and $T_2 = T_s$ with $\Delta T = 20\,{\rm K}$ and vice versa for the backward case.

\begin{figure}
	\centering
	\includegraphics[width = 0.45\textwidth]{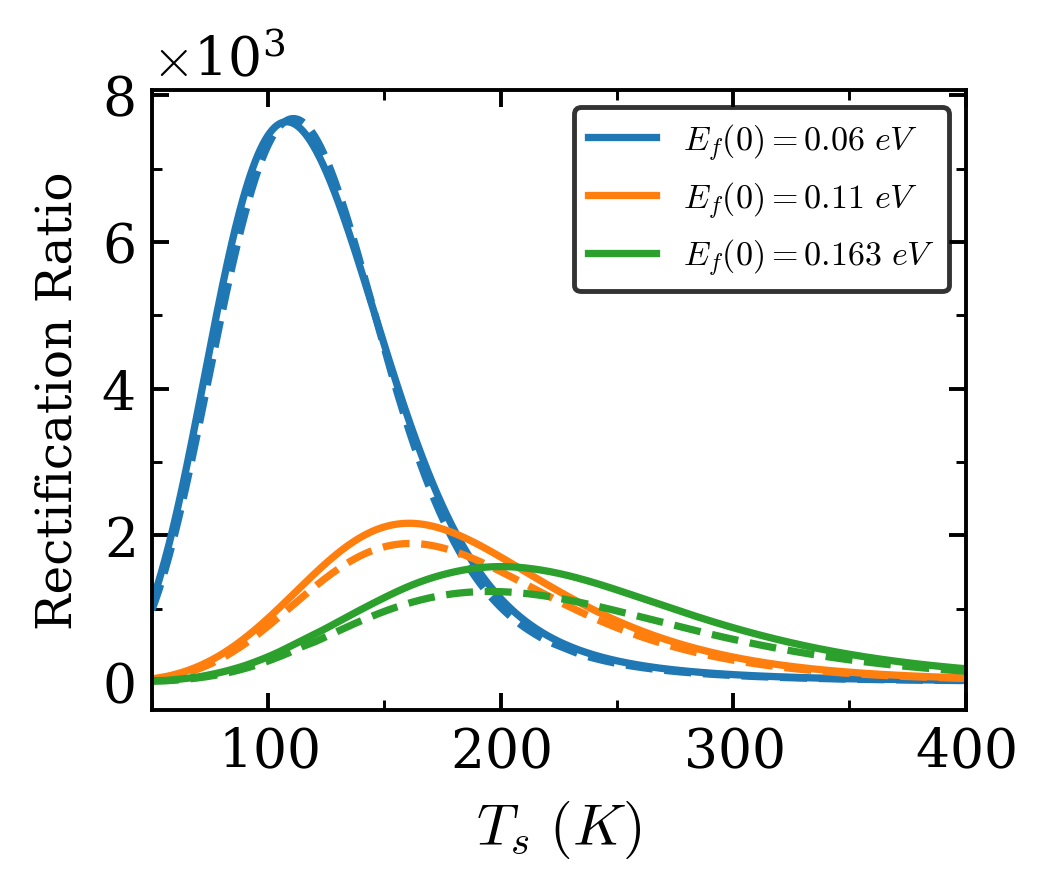}
	\caption{Rectification ratio $\eta$ from Eq.~(\ref{Eq:eta}) as function of temperature for diferent values of $E_F(0\,{\rm K})$ as tabulated in table~\ref{table:TabulatedWSMParams} for an interparticle distance of $d = 1000$nm using $\tau = 1000\,{\rm fs}$ (solid lines) and  $\tau = 450\,{\rm fs}$ (dashed lines). All other material parameters are the same as for the typical WSM discussed in the previous sections. 
	\label{Fig:FermiTemperature}}
\end{figure}

	In Fig.~\ref{Fig:FermiTemperature} we present the rectification ratio as function of the temperature $T$ for an interparticle distance of 1000nm using different values of the Fermi energy as found for the different WSMs tabulated in table~\ref{table:TabulatedWSMParams} and two different values of the relaxation time $\tau$. All the other parameters are the same as for the typical WSM as studied in the previous sections. It can be nicely seen that the maximum rectification moves to lower temperatures when the Fermi energy is decreased. This is due to the fact that by reducing the Fermi energy the surface modes are red-shifted. For $E_F(0\,{\rm K}) = 0.06\,{\rm eV}$ (as for Co$_2$MnGa) we have a maximal rectification ratio of about 8000 for $T_s = 108\,{\rm K}$ which is close to the nitrogen temperature. Furthermore, we can conclude from the results for different relaxation times that larger dissipation has the tendency to reduce the rectification ratio. Therefore our results indicate that WSMs with a small Fermi energy like Eu$_2$IrO$_7$ are good candidates for an experimental realization of large heat flux rectifications at small temperatures close to the nitrogen temperature of 77K. In general, WSMs with a small value of the Fermi energy are good rectifiers at low temperatures and WSMs with a relatively large value of the Fermi energy are good rectifiers at higher temperatures. We emphasize that by electrical gating, for instance, one can also actively modulate the Fermi level and therewith the rectification efficiency.

\begin{figure}
	\centering
	\includegraphics[width = 0.45\textwidth]{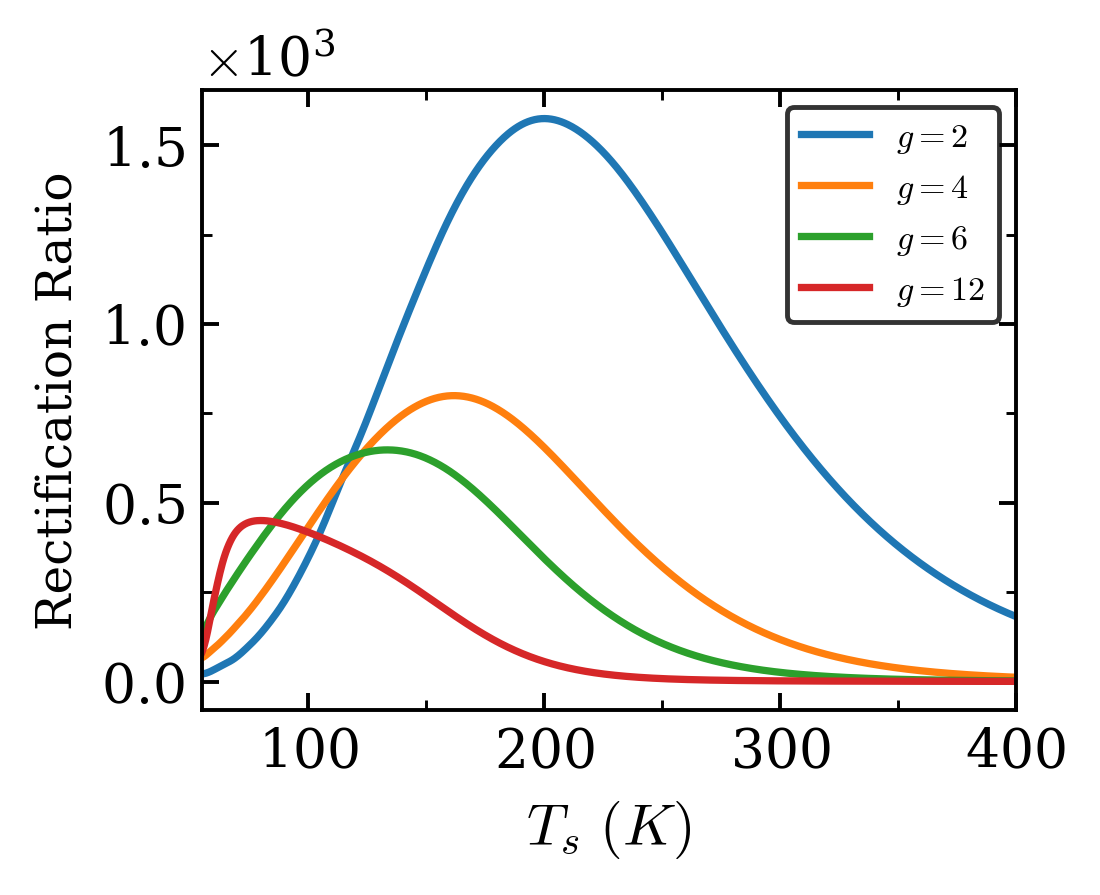}
	\caption{Rectification ratio $\eta$ from Eq.~(\ref{Eq:eta}) as function of temperature for diferent values of the number of Weyl nodes $g = 2, 4, 6, 12$ for an interparticle distance of $d = 1000$nm. All other material parameters are the same as for the typical WSM discussed in the previous sections. 
	\label{Fig:FermiTemperatureg}}
\end{figure}

\begin{figure}
	\centering
	\includegraphics[width = 0.45\textwidth]{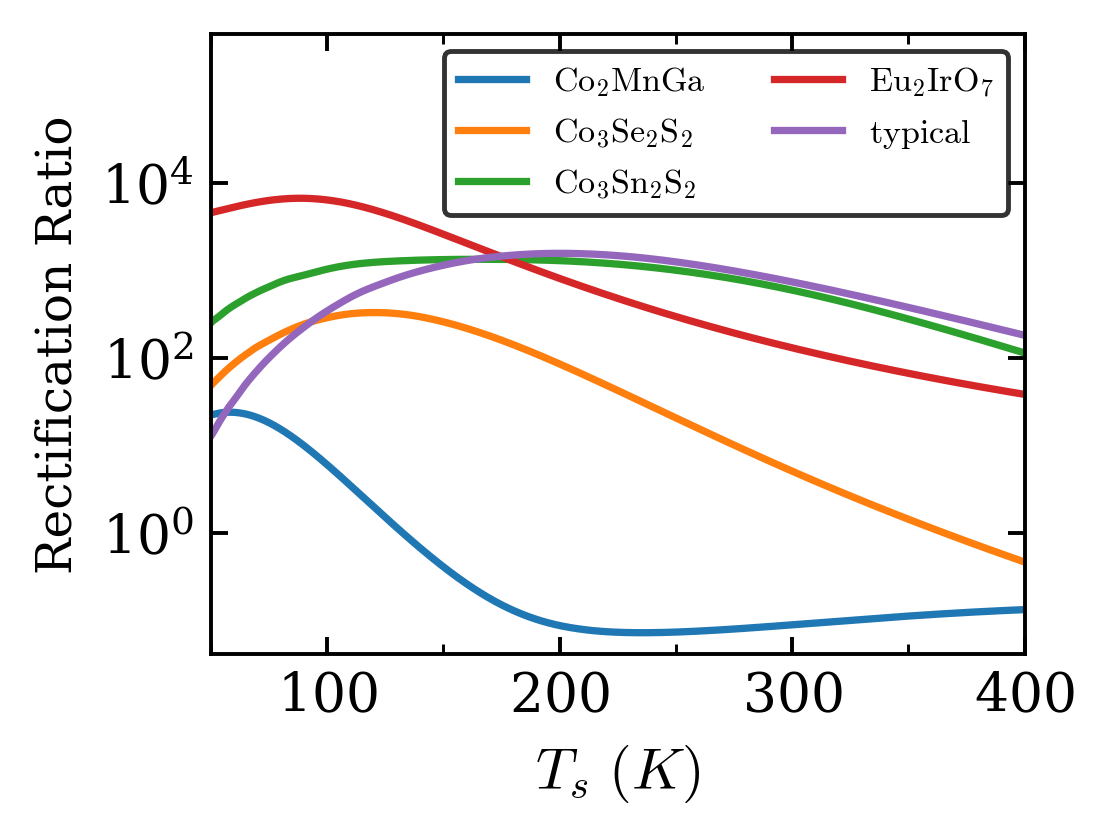}
	\caption{Rectification ratio $\eta$ from Eq.~(\ref{Eq:eta}) as function of temperature for all WSM tabulated in table~\ref{table:TabulatedWSMParams} with corresponding parameters and $d = 1000$nm.
	\label{Fig:FermiTemperatureTable}}
\end{figure}

	However, the rectification ratio depends on all parameters and in particular also on the number of Weyl nodes $g$ which has a strong influence on the permittivity. As can be seen in Fig.~\ref{Fig:FermiTemperatureg} we find that the more Weyl nodes we have the smaller the value of the rectification ratio will be. Hence, to have large heat flux rectification it seems preferable to use WSM with only two Weyl nodes. Of course, at the end one needs to find the best combination of all parameters including the material parameters $b$, $g$, $v_{\rm F}$, $\tau$, and $E_F(0\,{\rm K})$ and the interparticle distance $d$, the distance to the surface and the size of the nanoparticles. To have a better idea how existing WSMs could perform we show in Fig.~\ref{Fig:FermiTemperatureTable} the rectification ratio using the parameters for the WSMs tabulated in table~\ref{table:TabulatedWSMParams}. These curves suggests that WSMs with similar material properties as Eu$_2$IrO$_7$ are a very good choice for large rectification ratios at nitrogen temperatures and CO$_3$Sn$_2$S$_2$ as well as the typical WSMs studied in the previous sections are good rectifiers at somewhat higher temperatures around 200K.

%%%%%%%%%%%%%%%%%%%%%%%%%%%%%%%%%%%%%%%%%%%%%%%%%%%%%%%%%%%%%%%%%%%%%%%%%%%%%%%%
%
% Summary
%
%%%%%%%%%%%%%%%%%%%%%%%%%%%%%%%%%%%%%%%%%%%%%%%%%%%%%%%%%%%%%%%%%%%%%%%%%%%%%%%%

\section{Conclusion}

We have studied the rectification of thermal radiation between two WSM nanoparticles in the vicinity of a planar WSM sample in Voigt configuration. We have shown that the rectification direction and amplitude highly depends on the coupling of the localized nanoparticle resonances to the non-reciprocal surface modes of the planar substrate. We find extremely large rectification ratios in backward direction with values of $\eta \approx 5000-15000$ for the studied parameters. Furthermore, we find that the rectification ratios reported in Ref.~\cite{HuEtAl2023} of 2673 are incorrect as well as the explanation for the rectification ratios in terms of the permittivity ratio of the diagonal and non-diagonal elements of the permittivity tensor. We find for the same configuration values of 1502. Additionally, the rectification can rather be understood qualitatively and quantitatively by the properties of the surface modes and the coupling to either surface modes propagating in forward or backward direction. We highlight that for the parameters used in Ref.~\cite{HuEtAl2023} the rectification in forward direction with ratios of $\tilde{\eta} \approx 4-15$ for very small values of $b < 10^{8}\,{\rm m}^{-1}$ as well as the huge backward rectification $\eta \approx 8300$ for values of $b \approx 2\times10^{8},{\rm m}^{-1}$ have been overlooked in previous work in Ref.~\cite{HuEtAl2023}. This finding might open interesting possibilities for heat flux rectification for WSMs with small Weyl node separation. On the other hand, for values of $b = 2\times10^{9},{\rm m}^{-1}$ as found Co$_3$MnGa and Co$_2$MnGa we find a large rectification in backward direction only in the studied parameter regime. Even though the Weyl node separation can be modified electrically, magnetically, and opticially for certain WSM~\cite{TsaiEtAl2020,RayEtAl2020,LiEtAL2020} there are also other WSMs with with similar parameters as Co$_3$MnGa and Co$_2$MnGa but with smaller values of $b$ like Co$_3$Sn$_2$S$_2$, for instance. By turning the sample the directionality can be inversed but the coupling to the surface modes will be reduced so that a rectification in forward direction can be obtained. This rectification is much smaller than the rectification in backward direction but the rectification ratios up to about 300-900 which are still very substantial compared to the rectification ratio of 249 reported for InSb with applied magnetic fields~\cite{OttSpin2020}. Finally, our parameter analysis shows that the rectification ratio is very sensitive to the WSM parameters and for some parameters even very large rectification ratios larger than 15000 are in principle possible demonstrating the strong potential of WSM for heat flux rectification. Additionally we show that WSM with parameters such as Eu$_2$IrO$_7$ have a rather high rectification ratio of abut 6700 at temperatures close to nitrogen temperatures. For somewhat higher temperatures between 100K-177K we find that WSM with parameters such as Co$_3$Sn$_2$S$_2$ show still a large rectification ratio larger than 1000. These WSM are strong candidates for an experimental realisation of heat flux rectification.

%%%%%%%%%%%%%%%%%%%%%%%%%%%%%%%%%%%%%%%%%%%%%%%%%%%%%%%%%%%%%%%%%%%%%%%%%%%%%%%%
%
% Acknowledgement
%
%%%%%%%%%%%%%%%%%%%%%%%%%%%%%%%%%%%%%%%%%%%%%%%%%%%%%%%%%%%%%%%%%%%%%%%%%%%%%%%%

\section*{Acknowledgements}

The authors gratefully acknowledge financial support from the Niedersächsische Ministerium für Kultur und Wissenschaft (`DyNano') and disussions with Achim Kittel.

\end{document}